# Order and Information in the Patterns of Spinning Magnetic Micro-disks at the Air-water Interface

Wendong Wang[1,2†*], Gaurav Gardi[1†], Paolo Malgaretti[3], Vimal Kishore[1,4], Lyndon Koens[5], Donghoon Son[1,6], Hunter Gilbert[1,7], Zongyuan Wu[2], Palak Harwani[1,8], Eric Lauga[9], Christian Holm[10], Metin Sitti[1,11,12*]

[1] Physical Intelligence Department, Max Planck Institute for Intelligent Systems, 70569, Stuttgart, Germany
[2] University of Michigan – Shanghai Jiao Tong University Joint Institute, Shanghai Jiao Tong University, Dong Chuan Road 800, Minhang, Shanghai, 200240, China
[3] Helmholtz Institute Erlangen-Nürnberg for Renewable Energy (IEK-11), Forschungszentrum Jülich, Germany
[4] Department of Physics, Banaras Hindu University, Varanasi, 221005, India
[5] Department of Mathematics and Statistics, Macquarie University, Australia
[6] School of Mechanical Engineering, Pusan National University, Busan, 46241, Korea
[7] Department of Mechanical and Industrial Engineering, Louisiana State University, United States of America
[8] Department of Electronics and Electrical Communication Engineering, Indian Institute of Technology Kharagpur, India
[9] Department of Applied Mathematics and Theoretical Physics, University of Cambridge, United Kingdoms
[10] Institute for Computational Physics, University of Stuttgart, Allmandring 3, 70569, Stuttgart, Germany
[11] School of Medicine and School of Engineering, Koç University, 34450, Istanbul, Turkey
[12] Institute for Biomedical Engineering, ETH Zurich, 8092 Zurich, Switzerland
[†] These authors contributed equally.
* Correspondence to wendong.wang@sjtu.edu.cn, sitti@is.mpg.de

Abstract:

The application of the Shannon entropy to study the relationship between information and structures has yielded insights into molecular and material systems. However, the difficulty in directly observing and manipulating atoms and molecules hampers the ability of these systems to serve as model systems for further exploring the links between information and structures. Here, we use, as a model experimental system, hundreds of spinning magnetic micro-disks self-organizing at the air-water interface to generate various spatiotemporal patterns with varying degrees of orders. Using the neighbor distance as the information-bearing variable, we demonstrate the links among information, structure, and interactions. Most importantly, we establish a direct link between information and structure without using explicit knowledge of interactions. Finally, we show that the Shannon entropy by neighbor distances is a powerful observable in characterizing structural changes. Our findings are relevant for analyzing natural self-organizing systems and for designing collective robots.

## Introduction

The quest to seek the links between structure and information may be traced back to the idea of an "aperiodic crystal" as an information-carrying entity in living systems by Erwin Schrödinger (*1*), which portended the discovery of DNA (*2*, *3*). Almost in parallel, the Shannon entropy was introduced to quantify the amount of information in written texts in 1940s (*4*). Since then, its application in characterizing the structures of many systems, including organic molecules (*5*, *6*) and crystals (*7*–*10*), has yielded fruitful insights. For example, the replication and the operation of living systems requires an enormous amount of information, and the storage of this information necessitates molecules with very complex structures that are improbable to form under equilibrium conditions. This information-based argument on the molecular complexity suggests that the probability of life emerging under equilibrium is small (*6*), and therefore life must have emerged under non-equilibrium conditions (*11*). As another example, the application of the Shannon entropy in crystallography has led to the notion of chaotic crystallography and the creation of a continuous measure to quantify the degree of order/disorder in crystals (*7*). Similarly, the application of the Shannon entropy has also provided a precise quantitative answer to the question of which inorganic crystals are the most complex (*8*, *9*) and shown that the Shannon entropy contributes negatively to the thermodynamic configurational entropy of crystals (*10*). As a final example, the application of Shannon entropy in characterizing out-of-equilibrium systems has borrowed the notion of algorithmic complexity pioneered by Kolmogorov and Chaitin (*12*, *13*) and led to the usage of an information measure based on lossless data compression to quantify hidden order in simulated model systems such as absorbing state models and active Brownian particles (*14*).

Although the application of the Shannon entropy has yielded valuable insights for the systems mentioned above, much remains to be learned about the relationship between the abstract notion of information and its concrete manifestation in a structure. Those molecular and crystal systems have limitations as model systems because it is difficult to manipulate and directly observe the mutual interactions of the atoms and the molecules. Although simulations on these systems have provided valuable insights, a combined approach based on experiments, theory, and simulation to investigate one model system in detail could provide an archetypical case study that sheds light on other systems. Indeed, an ideal model system should consist of trackable objects whose mutual interactions are tunable and could be modelled and analyzed theoretically and numerically.

A self-organizing system at the micrometer scale and above could be an ideal model system to study the relation between information and structure. The self-organization in many natural (*15*–*19*) and artificial (*20*–*26*) collective systems display spatiotemporal patterns over the length scales of micrometers to meters and over the time scales of milliseconds to seconds. One distinguishing feature of these patterns is their spatiotemporal order. In particular, torque-driven spinning particles such as millimeter-sized disks (*27*, *28*), magnetic colloids (*29*), micro-rafts (*30*), and biological systems such as spinning bacteria (*31*) and ATP synthase (*32*) often display two dimensional (2D) hexagonally-ordered patterns. The constituents of most of these 2D patterns can be directly observed and tracked by conventional light microscopy. However, most of these microscopic systems form only a few patterns (typically two, ordered and disordered), so their patterns lack the diversity necessary for studying how information changes as patterns vary. This lack of diversity could be attributed to the lack of tunability in the mutual interactions among microscopic constituents.

Here we use the diverse spatiotemporal patterns in the self-organization of hundreds of spinning micro-disks trapped at the air-water interface as a model system to demonstrate the relation between the information and the order of the patterns. We show how careful tuning of local pairwise interactions and local symmetries produces a wide range of global patterns with varying degrees of order. We apply the formulation of the Shannon entropy to the graphs corresponding to the patterns (*6*, *33*, *34*) and show how neighbor distances (defined through Voronoi tessellation) arise naturally as the information-bearing variable for calculating the Shannon entropy. Next, we use the distribution of neighbor distances to reproduce *in silico* patterns characterized by the same orientational orders, thereby highlighting a direct link between information and order. Finally, we show that the entropy by neighbor distances is a more powerful observable for detecting both spatial and temporal changes of the patterns than the orientational order parameters.

## Results

### Balancing attractive and the repulsive interactions

To begin, we consider the balance of attractive and repulsive forces in local interactions. The mutual interactions between a pair of spinning magnetic micro-disks at the air-water interface include magnetic, capillary, and hydrodynamic interactions (**Fig. 1A**). In our current setup, the hydrodynamic lift force (*27*, *28*) and the angle-averaged capillary force (*30*) produce the mutual repulsion, and the effective magnetic interaction between two synchronously-rotating magnetic dipoles produces the mutual attraction (*35*). We use a custom-made two-axis Helmholtz coil to generate a rotating uniform magnetic field (**fig. S1**) and sputter thin films of cobalt on micro-disks to generate an in-plane magnetic dipole for each micro-disk. Under the rotating magnetic field and above a certain critical threshold rotation speed, individual micro-disks rotate around their own axes, and we approximate their mutual interactions with the angle-averaged interactions.

We first consider magnetic dipole-dipole interactions. It is solely responsible for mutual attraction between two micro-disks. Quantitatively, the angle-averaged magnetic dipole-dipole attraction is expressed as $F_{mag} = -\frac{3}{4}\pi^2 \mu_0 \rho_m^2/(d/R)^4$, where $\mu_0$ is vacuum permeability; $d$ is the center-center distance; $R$ is the radius of the disk; $\rho_m$ is the magnetic moment per unit area and depends on the thickness of the sputtered cobalt thin film. For a 500-nm-thick film, we have $\rho_m \approx 0.1$ A and $F_{mag} \approx -3\pi^3/(d/R)^4 \cdot 1\text{nN}$. At a fixed $d/R \sim 2-3$ , $F_{mag}$ is a constant and is on the order of 1 nN (see Supplementary notes on the scaling relations for more details).

Next, we choose parameters of the micro-disks such that the capillary and hydrodynamic interactions are of the same order of magnitude as the magnetic interactions (~ 1 nN) to strike a balance between the attractive and repulsive interactions. The capillary interaction is due to the cosinuisoidal edge profiles (**fig. S2**) around the micro-disks (*36*, *37*), and the hydrodynamic lift force is due to the fluid inertia at finite Reynolds number (*27*, *28*). The capillary interactions dominate in the near field ($d < 2.5R$), whereas the hydrodynamic interactions' relative influence increases as $d$ increases. Both forces depend on radius $R$, but the capillary force can be independently adjusted by the amplitude and the arc angle of the cosinusoidal profiles. Quantitatively, by decomposing the edge profiles into a series of Fourier modes in bipolar coordinates (*36*, *37*), we find simple numerical relations between the angle-averaged capillary force $F_{cap}$ [N] and $R$ [m] at fixed $d/R$'s. With the amplitude being 2 μm and the arc angle being 30° (**fig. S2A**), $F_{cap} \sim 10^{-13}\mathrm{N} \cdot \mathrm{m} \cdot R^{-1}$ for $d \sim 2.5R$. Therefore, for $R \sim 10^{-4} m$, $F_{cap}$ is ~ 1 nN (see Supplementary notes on the scaling relations for more details).

On the other hand, the hydrodynamic lift force follows a simple scaling relation: $F_{hydro} \approx \rho\omega^2 R^7/d^3$, where $\rho$ is the fluid's density, and $\omega$ is the spin speed. Using Reynolds number $\mathrm{Re} = \omega R^2/\nu \sim 1$, where $\nu$ is the fluid's kinematic viscosity, we recast the expression as $F_{hydro} \approx \rho\nu^2\mathrm{Re}^2/(d/R)^3 = \mathrm{Re}^2/(d/R)^3 \cdot 1\ \mathrm{nN}$, where $\rho\nu^2$ [N] depends only on the properties of the water and is ~ 1 nN. Re can be adjusted either by changing the radius $R$ during fabrication or by varying the rotation speed $\omega$ during experiments. Because our coil system can produce a uniform rotating magnetic field of ~70 revolutions per second (rps) for a few minutes without overheating, we have chosen $R$ = 150 μm so that Re can reach ~ 10 in our experiments.

This system differs from the previous reports (*27*, *28*, *30*) in which a global magnetic potential provides the effective attraction towards the center of the potential. Because all the interactions between micro-disks can be considered as pairwise interactions in our current setup, the system of many micro-disks could possess a richer collection of patterns. Moreover, we symmetrically position 4 – 6 cosinusoidal profiles around the edge of a micro-disk to produce different local symmetry in the deformation of the air-water interface around the micro-disk. The variation in the local symmetry does not affect the behaviors of spinning micro-disks as long as they can spin freely around their own axes. It is only when they start to attach at low spin speeds ($\omega \leq 10$ rps) that the local symmetry shows its effect. At first, we focus on micro-disks with 6-fold symmetry.

**Regions of pairwise interactions relate to different patterns of many micro-disks**

Systematic study of pairwise interactions reveals three distinct regions (**Figs. 1B, C** and **Movie S1**): the two micro-disks (I) attach to each other, (II) orbit around each other, and (III) move away from each other. Region (I) and (II) have been observed previously in the case of a global magnetic potential (*30*), and the transition from (II) to (I) is due to the increased oscillation around mean steady-state separation distance as the rotation speed decreases and the capillary torque locking the alignment of the micro-disks (*36*). Region (III) is new and is due to the increase in the hydrodynamic lift force as spin speeds increase, as confirmed by a two dimensional (2D) numerical pairwise model constructed with experimental values and without fitting parameters. The numerical result (**fig. S3**) shows that as the spin speed increases above 22 rps, the increasing hydrodynamic repulsion makes the sum of forces repulsive at all distances, thereby decoupling the pair of orbiting micro-disks.

Systematic study of the self-organization of hundreds of micro-disks reveals many visually distinct patterns. We first focus on patterns that appear at the spin speeds corresponding to the regions (II) and (III) of the pairwise interactions (**Fig. 1D** and **Movies S2 - S3**). At the spin speeds of the region (II), the patterns of many micro-disks appear disordered, whereas, at the spin speeds of the region (III), the patterns show hexagonal order. The appearance of the hexagonal order motivates the use of hexatic order parameter $\psi_6$ (see Materials and Methods section on the calculation of order parameters and **figs. S4A-B** for details) to quantify the orientational order (*38*). Specifically, we calculate an averaged norm of the hexatic order parameters $<|\psi_6|>_{N,t}$ , where the subscripts $N$ and $t$ denote the number average within one frame and time average over many frames, respectively. We find a sharp transition of $<|\psi_6|>_{N,t}$ at around 23 rps (**Fig. 1E**), which coincides with the pairwise transition from the region (II) to (III). Moreover, by assuming each micro-disk interacting with the rest of the micro-disks through pairwise interactions and with the physical boundary, we obtain a 2D numerical model of many micro-disks that also captures the transition of $<|\psi_6|>_{N,t}$ at around 23 rps (**figs. S4C–E** and **Movie S4,** see Materials and Methods section on Model for many-disk interactions for details).

## Hamiltonian approach

This close correspondence between the pairwise transition and the many-disk transition from the region (II) to (III) motivates us to seek a more fundamental link between them. Because all the interactions between two micro-disks can be assumed to be of pairwise nature (*i.e.*, not produced from a global potential), we can construct an effective Hamiltonian as a function of the separation distance between a pair of neighboring micro-disks. Neighbors are defined by Voronoi tessellation. Specifically, we construct the one-dimensional (1D) effective Hamiltonian of pairwise interactions $H(d)$ as a function of the separation distance $d$ between the pair of micro-disks. We introduce a mean field energy term $E_{mf}$ to account for all the interactions of the pair with the rest micro-disks and with the physical boundary. Therefore, the Hamiltonian can be written as

$$H(d) = E_{magdp}(d) + E_{cap}(d) + E_{hydro}(d) + \Gamma \cdot E_{mf}(d), \qquad (1)$$

where $d$ is the pairwise distance; $E_{\mathrm{magdp}}$ is the angle-averaged magnetic dipole-dipole energy; $E_{\mathrm{cap}}$ is the angle-averaged capillary energy; $E_{hydro}$ represents the effective energy associated with the hydrodynamic interaction and is calculated from the integration of the hydrodynamic lift force; $E_{mf}$ represents the mean field energy term. More specifically, $E_{mf}$ is calculated as the mean of interactions by all other micro-disks on the pair under consideration, with the assumption of a uniform area density of other micro-disks. Finally, $\Gamma$ is a fitting parameter that accounts for all the discrepancies because of the simplifications used in order to construct the closed-form expression for the $E_{bd}$ (see **fig. S5A** and Supplementary Note on the Hamiltonian approach for more details). We found that $\Gamma$ is 10 for all spin speeds.

From this 1D effective Hamiltonian $H(d)$, we calculate the distribution of pairwise distances, assuming (*39*) that they are distributed according to the Boltzmann factor $p(d) \propto \exp(-\beta H(d))$, where $\beta$ is an additional fitting parameter. The intuition behind this assumption is that in region II and III, the angle-dependent capillary and magnetic interactions creates a time-varying attraction/repulsion between a pair of micro-disks, which generates an effective fluctuation along the radial direction of the micro-disk. As a result, for the degree of freedom along the radial direction, the effective fluctuation enables the micro-disks to explore the full range of the 1D energy landscape. The calculated distributions are fitted with the experimental distributions of neighbor distances (**fig. 2A**) to obtain the fitted values of $\beta$. Because in equilibrium systems $1/\beta$ is the thermal energy, we compare it with the variance of the pairwise distance (**fig. 2B**). The variance of the neighbor distances is calculated as $\sigma_{NDist} = \sum_{\forall d} < [\, d - < d >]^2 >$ and correlates well with $1/\beta$, so we regard $1/\beta$ as the effective energy governing the fluctuations of the neighbor distance.

To compare the fitted probability distribution with the experimental ones across all spin speeds, it is useful to have a single-valued observable. To this end, we calculate the Shannon entropy associated with the probability distribution of neighbor distances as

$$H_{NDist} = -\sum_i p_i \log_2(p_i), \tag{2}$$

where $p_i = X_i/X$ is the probability of a neighbor distance that falls within a distance interval (a bin) labelled by index $i$; $X$ is the total count of all neighbor distances of all micro-disks, and $X_i$ is the count of the neighbor distances in bin $i$. We have found that the choice of bin size in the range of $0.1 - 0.8R$ does not affect the results, so we have chosen $0.5R$ as the bin size (See **fig. S5B** for more details). For steady states, $H_{NDist}$ is calculated from the distribution of all the neighbor distances for the whole duration of observation (See Supplementary Note on the Hamiltonian approach for more details). From the information-theoretic perspective, $H_{NDist}$ represents the average information content of an event that measures the distance between a random pair of neighboring micro-disks (*40*). Intuitively, the smaller the value of $H_{NDist}$, the narrower the distribution of the neighbor distances. (To illustrate the idea of information content, consider rolling a die or flipping a coin: the information content of casting a die once is $-\log_2(1/6) = \log_2 6$, and the information content of flipping a coin once is $-\log_2(1/2) = \log_2 2$. Therefore, the Shannon entropy, or the average information content, of a single die-casting is higher than a single coin-flipping.)

**Fig. 2C** shows the good agreement between the Shannon entropies calculated from the experimental distributions and those calculated from the fitted probability distributions. This good agreement not only suggests that the terms included in the effective Hamiltonian are enough to explain the variety of patterns but also that $H_{NDist}$ can characterize the structural changes in the patterns. Indeed, the drop of $H_{NDist}$ around ~20 rps captures the transition between region (II) and (III) (**Fig. 1E**). Moreover, the increase of $H_{NDist}$ from 11 rps to 15 rps suggests an additional transition. This transition is not clearly distinguishable by $<|\psi_6|>_{N,t}$ (**Fig. 1E**), but the large change in $H_{NDist}$ suggests that the patterns at 11 – 12 rps are qualitatively different from the patterns at 15 – 20 rps. Indeed, we observe that the patterns at 11 –12 rps consist of a densely packed core surrounded by clusters of single or few micro-disks as if they were a mixture of condensed and dispersed phases. Additional experiments (to be reported elsewhere) indicate that it is possible to obtain a pure condensed phase, in which micro-disks tightly packed but still able to rotate freely relative to each other.

To quantify the information embedded in the patterns, we compare the experimental distributions of neighbour distances with a reference distribution generated from randomly positioned non-overlapping micro-disks (**fig. S6**). The Kullback – Leibler divergence between the experimental distributions and the reference distribution represents the extra information embedded in the experimental patterns. Thus, the plot of this divergence as a function of the rotation speed of the external magnetic field (**fig. S6C**) is almost a mirror image of the corresponding $H_{NDist}$ plot (**Fig. 2C**): the more ordered the pattern is, the more the pattern deviates from a random pattern, and the more extra information the pattern contains as compared with a random pattern. This perspective is similar to the maximum entropy principle advocated by Jaynes (*41*): the addition of new information changes the distribution of the random variable that embeds the information. This line of thought leads us to explore a direct link between structure and information, as elaborated in the section on Monte Carlo simulation below.

## Monte Carlo approach

The preceding analysis suggests that the pairwise interactions serve as an intermediate bridge between the order and the information of the patterns. We have seen that a simple extension of the 2D numerical pairwise model to many-disks reproduces the change of the order from the region (II) to (III) (**Figs. 1C, 1E** and **fig. S4**) and that the construction of an effective 1D Hamiltonian based on the pairwise interactions reproduces the change in the Shannon entropies by neighbor distances from the region (II) to (III) (**Fig. 2C**). Now we ask: *are there any direct links between the order of a pattern and the distributions of its neighbor distances without resorting to either the numerical pairwise model or the effective Hamiltonian?*

To address the above question, we perform Monte Carlo simulations to see if it is possible to recreate the spatial order observed in experiments from the information contained in the probability distribution of the neighbor distances. Specifically, we start from initially randomly distributed micro-disks and accept (or reject) the move of a micro-disk if the move decreases (or increases) the Kullback–Leibler divergence (KLD) (*42*, *43*)

$$D_{KL}(P(d)\,||\,Q(d)) = \sum p(d)\ln\frac{P(d)}{Q(d)}, \tag{3}$$

where $P(d)$ is the simulated distribution, and $Q(d)$ is the experimental distribution (see Supplementary note on the Monte Carlo simulation for details). Intuitively, KLD quantifies how different the two distributions are. We use the four representative patterns (**Fig. 1E**) from the experiments. With only local information embedded in the distributions of neighbor distances, we are able to recreate all four representative patterns with orders that are comparable with the experimental values (**Fig. 3, Table S1**). Moreover, the simulated patterns also show the marginal distributions of $x$ and $y$ coordinates that match the experimental values. The radial distribution functions (**Fig. 3E**) also show good match between experiments and Monte Carlo simulations. These agreements further validate the choice of the neighbor distances as the information-bearing variable.

**Extending Shannon entropy by neighbor distances to patterns formed by micro-disks with different local symmetries**

Last, we extend our analysis of information and order to the patterns of micro-disks with different local symmetries. Because the pairwise interaction in the region (II) and region (III) can be treated in an angle-averaged manner, the resulting patterns do not differ for micro-disks of different symmetries. It is only when the micro-disks start to attach to each other to form 2D tiles that the local symmetries of the micro-disks start to affect the global patterns. Therefore, in the following tiling experiments, we gradually decrease the spin speeds of the magnetic field $\Omega$ and observe the patterns formed by hundreds of micro-disks with 4, 5, or 6 cosinusoidal profiles symmetrically distributed along the edge of the micro-disks.

For micro-disks with 6-fold symmetry (**Fig. 4A and Movie S5**), the patterns include hexagonally ordered pattern at $\Omega > 22$ rps, disordered patterns at $\Omega \sim 22 - 10$ rps, and clusters at $\Omega \sim 10 - 1$ rps and a crystal-like pattern for $\Omega < 1$ rps. We found that mixing low magnetic field strengths (0.5 mT) at $\Omega = 0.25$ rps with short bursts of high field strength (3 mT) at $\Omega \geq 1$ rps produces an effect similar to annealing in crystal growth. The 6-fold symmetry of micro-disks generates a crystal-like pattern with local 6-fold symmetry, so $\psi_6$ can be used to track the change in the structural order of the entire tiling process (**Fig. 4B**). Significantly, entropy by neighbor distances $H_{NDist}$ also displays high sensitivity in detecting subtle changes of structural order throughout the process: a drop in $<|\psi_6|>_N$ always corresponds to a rise in $H_{NDist}$. Indeed, the two observables are almost completely anti-correlated, with a Pearson correlation coefficient of -0.99 (**fig. S7A**). Besides neighbor distances, the statistics of two other local variables, neighbor counts (*44*) and local densities (local volumes) (*45*), have been proposed to characterize the structures of packing in 2D. However, the Shannon entropy calculated based on the distribution of neither neighbor counts nor local densities shows a good correlation with $<|\psi_6|>_N$ (**fig. S7B – C**), thus highlighting the unique effectiveness of $H_{NDist}$ in distinguishing the order in the patterns.

For micro-disks with 4-fold symmetry (**Fig. 4C** and **Movie S6**), however, the tiling process started with a hexagonally ordered pattern but ended with a crystal-like pattern with local 4-fold rotational symmetry. As a result, the quantification of the order requires two types of order parameters: the high hexagonal order at the beginning of the process is identified by the large value of $<|\psi_6|>_N$, and the high tetragonal order at the end of the process is identified by the relatively large value of $<|\psi_4|>_N$. However, these two ordered patterns with different local symmetries both show small values of $H_{NDist}$, suggesting that $H_{NDist}$ is a more universal observable for the identification of order than $<|\psi_6|>_N$ or $<|\psi_4|>_N$ (**Fig. 4D**). Even for micro-disks with 5-fold symmetry, which are only capable of forming "amorphous" tiling, $H_{\mathrm{NDist}}$ is most sensitive to the periodicity in the Mix part of the tiling process (**Figs. 4E – 4F** and **Movie S6**): the Fourier spectrum of $H_{NDist}$ shows the strongest signal-to-noise ratios with multiple clear high order peaks than either $<|\psi_6|>_N$ or $<|\psi_5|>_N$ (**fig. S8**), clearly showing the temporal structure of the patterns. Moreover, because $H_{NDist}$ is not symmetry-specific, it can be used to compare the degree of orders in the tiling of different symmetries: micro-disks with 6-fold symmetry produces the lowest $H_{NDist}$, because the hexagonal packing tolerates misalignment better than square packing (**fig. S9**).

Via these tiling experiments, we demonstrated the effectiveness of the Shannon entropy by neighbor distances $H_{NDist}$ in characterizing the structural orders and in detecting subtle structural changes. Compared with the Shannon entropies of other quantities like neighbor counts (*44*) and local densities (local volumes) (*45*), $H_{NDist}$ is particularly effective in distinguishing different patterns. We speculate that the particular effectiveness of $H_{Ndist}$ is due to the intimate relations between the physics of the system and the neighbor distance. From this perspective, we expect that when used as feedback, $H_{NDist}$ could be helpful for the control of robotic swarms (*46*, *47*) where a change in the internal driving force or the external boundary results in a change of global patterns.

## Discussion

Our results show close relations among information, structures, and interactions (**Fig. 5**). We demonstrate direct links between each pair of them via different approaches. First, we reproduced the experimental patterns (structures) via the 2D numerical model based on pairwise interactions and interactions with the boundary. This approach connects the structures with the detailed interactions. Second, we reproduced the distributions of neighbor distances using an effective 1D Hamiltonian based on pairwise interactions and a mean field energy term. This approach connects the information to the interactions. Third, we reproduced the experimental patterns via Monte Carlo simulations using the distributions of neighbor distances. This approach connects the structure with the information without using any explicit knowledge of the interactions of the system. Therefore, it is particularly useful for systems where it is hard or impossible to determine the effective interactions among the constituents. This situation is likely to occur for systems made of computing units, such as biological cells, animals, humans, and robots.

Even though our current system is a planar 2D system, the method of calculating $H_{NDist}$ could extend to three-dimensional (3D) system. The increased degrees of freedom in 3D suggests that it is in principle possible for multiple configurations to satisfy one particular neighbour distribution. For example, both face-centered cubic packing (FCC) and hexagonal close packing (HCP) have 12 nearest neighbours of distance $2R$, so their $H_{NDist}$ will be the same. Nevertheless, like the case in 2D where $H_{NDist}$ would be able to distinguish between imperfectly packed squares and hexagons, $H_{NDist}$ should be able to distinguish between imperfectly packed simple cubic lattice and HCP or FCC lattice because the body diagonal lattice point of simple cubic lattice is also a neighbour via Voronoi tessellation. The other challenge in the application of $H_{NDist}$ in 3D is perhaps the difficulty in measuring accurately the positions of all the particles and tracking them over time. Current available techniques for tracking particles in 3D at the microscopic scale include spinning disk confocal microscope, light sheet fluorescent microscope (*48*), and digital holographic microscope (*49*). If the positions of particles can be measured and tracked over time, then a Voronoi tessellation in 3D could be used to define neighbours, and the calculation of $H_{NDist}$ could proceed as usual. Therefore, for these 3D microscopic cases, the method based on $H_{NDist}$ will work. Another possible scenario of the successful application of the $H_{NDist}$ method in 3D is a robotic swarm in which individual robots can sense the distances to their neighbours without knowing the precise global coordinates of the neighbours. Data gathered from these robots could feed directly into the calculation of $H_{NDist}$ and enable subsequent analysis of self-organized patterns.

A limit of our experimental system is the necessity of a physical boundary in the formation of patterns of many micro-disks. Indeed, we have chosen a square-shaped boundary specifically to highlight the two different patterns in the region (III). If used constructively, however, this need for a physical boundary could be useful for designing experiments that explore the interactions between the self-organizing patterns and their global environments.

A possible extension of Monte Carlo simulation method is to replicate the pattern of a single frame. The target distributions in Fig. 3 are based on the neighbour distance data collected across multiple frames and are in this sense the steady-state averages. If, however, we collect the neighbour distances from only one frame to calculate the distribution and try to reproduce the pattern using this distribution as the target, we could reproduce the order in transient states, such as the ones shown in the Fig. 4. Indeed, we have attempted at reproducing a few transient patterns using the distributions of neighbour distances from a single frame. The results are summarized in **fig. S10** and **Table S2**. The most ordered pattern (0.25 rps after Mix) is most difficult to simulate, probably because it possesses both long-range and short-range orders. Thus, the simulations often get trapped into many local minima while trying to reach the correct pattern. Therefore, additional procedures like equivalent of annealing (or mix in our experiments) may be required to reach global minimum.

Finally, we envision that our experimental system could be used for testing hypotheses such as non-equilibrium pressure (*50*) and non-ergodicity in hydrodynamic self-organization (*51*). In the long term, this system could be used to design collective robotic systems to process information and perform computations (*52*).

## Materials and Methods

### Preparation and characterization of the micro-disks

Micro-disks were designed in Rhinoceros 3D with the aid of the Grasshopper plug-in. They were fabricated on Nanoscribe Photonic Professional GT with a 25x objective and with IP-S photoresist in the dip-in mode. The slicing distance was set to be adaptive from a minimum of 0.5 μm to a maximum of 3 μm. The hatching distance was 0.3 μm; the hatching angle 45°; hatching angle offset 72°. The number of contours was 3.

Thin films of ~500 nm cobalt and ~60 nm gold were sputtered onto the micro-disks using Kurt J. Lesker NANO 36. The base vacuum pressure before the sputtering was $<5 \times 10^{-7}$ Torr. Cobalt was sputtered at 100 W and under a sputtering pressure of $\sim 4.2 \times 10^{-3}$ Torr; gold was sputtered at 40 W and under a sputtering pressure of $\sim 2.7 \times 10^{-3}$ Torr. The gold layer is to protect the cobalt layer from oxidation. The sputtering procedure could be finished within one day.

We increased the diameter of micro-disks from 100 μm to 300 μm and increased the thickness of the cobalt layer from 50 nm to 500 nm, thereby increasing the magnetic moment ~100-fold compared with our previous reports. Consequently, the angle-averaged magnetic dipole force dominates in the far field ($d > \sim 100$ μm), whereas the angle-averaged capillary force dominates in the near field ($d < \sim 30$ μm). In the intermediate distances, the balance between the two main pairwise forces creates a coupled steady state: two micro-disk orbit around each other at medium rotation speeds ($\Omega = \sim 10 - 20$ revolution per second (rps)).

Scanning electron microscope (SEM) images of the micro-disks were taken on EO Scan Vega XL at 20 kV. Laser scanning confocal microscope images were taken on Keyence VK-X200 series with a 20x objective. The optical microscope images were taken on Zeiss Discovery V12 using Basler camera acA1300-200uc. The magnetic hysteresis curves of 500 nm cobalt film sputtered on a 30-mm-diameter coverslip was measured on MicroSense Vibrating Sample Magnetometer (VSM) EZ9. Digital holographic microscopy (DHM) images were recorded and analyzed on Lyncée Tec reflection R2200 with a 5x or 10x objective.

### Video acquisition

Experimental videos were recorded using Basler acA2500-60uc or Phantom Miro Lab140. The cameras were mounted on Leica manual zoom microscope Z16 APO. LED light source SugarCUBE Ultra illuminator was connected to a ring light guide (0.83'' ID, Edmund Optics #54-176), which was fixed onto the coil frame using a 3D-printed adapter.

The experimental videos were analyzed with a custom Python code using the OpenCV library. For pairwise data, the positions and the orientations of the micro-disks were extracted to calculate edge-edge distances and angular orbiting speeds. For many-disk data, the positions of micro-disks were extracted. Voronoi diagrams were constructed to identify neighbors and to calculate $\psi 6$, HNDist, and other parameters that characterize structural orders and information content.

### Fabrication and calibration of the custom electromagnetic coil systems

The custom-built Helmholtz coil system to generate uniform magnetic fields in the XY plane consists of two 5-cm-radius *x*-coils and two 8-cm-radius *y*-coils. The enamelled copper wire is 1.41 mm in diameter. The frames of the coil system were designed in Solidworks and 3D printed by Stratasys Fortus 450mc. The material of the coil frame is Ultem 1010, which has a high heat deflection temperature of 216 °C. Each coil was driven by an independent motor driver acting as a current controller (Maxon ESCON 70/10). The power for the current controllers was supplied by Mean Well, SDR – 960 – 48 (48 VDC at 20 A). The four motor drivers were connected to the analogue output channels of a National Instruments USB-6363, which was controlled by a LabView program on a PC. The dynamic performance of the current controllers was tuned manually in the vendor's software Maxon Studio, and the gain and integration time constants were adjusted so that the commanded currents were able to track signals up to 100 Hz without noticeable roll-off in magnitude or phase delays.

Each coil was independently calibrated by measuring the B-field in five locations in the workspace. The mapping was automated using a three-axis stage made of three linear stages (LTS300 Thorlabs). The measured B-field was used to calculate the current-to-B-field matrix Inversion of the $M_{IB}$ gives the B-field-to-current matrix $M_{BI}$.

$$M_{IB}\begin{pmatrix} I_1 \\ I_2 \\ I_3 \\ I_4 \end{pmatrix} = \begin{pmatrix} B_x \\ B_y \\ dB_x/dx \\ dB_y/dy \end{pmatrix}, \tag{1}$$

$$M_{BI} = M_{IB}^{-1}, \tag{2}$$

$$M_{BI}\begin{pmatrix} B_x \\ B_y \\ dB_x/dx \\ dB_y/dy \end{pmatrix} = \begin{pmatrix} I_1 \\ I_2 \\ I_3 \\ I_4 \end{pmatrix}. \tag{3}$$

**Simulation methods**

The capillary force and torque for edge-edge distances below 50 μm were simulated using Surface Evolver 2.7. A circle of 1mm in diameter was used as the outer boundary, and the two micro-disks were positioned along the x-axis and separated by an edge-edge distance from 1 μm to 50 μm. The orientations of the micro-disks were kept equal and varied from 0° to 60°. Total surface energy was obtained as a function of the edge-edge distance and the orientation of the micro-disks. The capillary force was obtained as the negative of the derivative of the energy over distance. The capillary torque was obtained as the negative of the derivative of the energy over the orientation angle.

The capillary force and torque for edge-edge distances above 40 μm were computed according to equations in the section on Capillary force and torque calculation in Matlab. The simulation for pairwise interactions and the collective phases of many disks were performed according to equations in the sections on the model for pairwise interactions and model for many-disks interactions in Python. In all simulations, the direction of the magnetic dipole is assumed to coincide with one of the six peaks of the cosinusoidal edge profiles. The angle between the direction of the magnetic dipole and the x-axis is considered as the orientation of the micro-disk.

In the pairwise simulations, the initial edge-edge distance of the two micro-disks was set to be 100 μm, and the initial orientation angles of the two micro-disks were set to be 0. The time step is 1 ms, and the total time varies between 2 – 50 s. The analysis of steady states was based on the last 2 s of simulation data. The integration is solved using the Explicit Runge-Kutta method of order 5(4) in the SciPy integration and ODEs library. We observe that a steady state was usually reached within 1 s.

In the simulations of collective patterns, the initial positions of the disks were aligned along a spiral on a square lattice. The center of the spiral is the center of the arena. The spacing between micro-disks is 100 μm. The time step is 1 ms, and the total time is 10 s. The integration is solved using the Explicit Runge-Kutta method of order 5(4) in the SciPy integration and ODEs library. We observe that steady states were reached after 6 – 7 s.

**Detailed experimental protocols**

Pairwise experiments (**Figs. 1B- 1C**) were performed in the arena of 8 mm diameter shown in **fig. S1D**. The air-water interface was kept flat by adjusting the amount of water. Videos were recorded in two sequences, one for each type of transitions: (1) Ω = 10 rps – 22 rps (decoupling transition), in steps of 1 rps (red curve in **Fig. 1C**), (2) Ω = 21 rps – 6 rps (assembling transition), in steps of 1 rps, (black curve in **Fig. 1C**). The field strength was 10 mT for all sequences. There was a gap of about 60 s between two rotation speeds to allow the micro-disks to reach steady states. 2 s of data were recorded for each rotation speed. The magnification of the zoom lens was 2.5x. We also performed pairwise experiments at other magnetic field strengths (1 mT, 5 mT, and 14 mT). These data will be reported separately.

Experiments with 218 micro-disks (**Figs. 1D – 1E**) were performed in the square arena with an edge length of 15 mm, shown in **fig. S1E**. For both flat and concave air-water interfaces, videos of 1 s were recorded for Ω = 70 rps – 10 rps in steps of 1 rps, and then videos of 20 s were recorded for Ω = 70 rps – 10 rps in steps of 10 rps. There was a gap of at least 60 s between two video recordings to allow the micro-disks to reach steady states. The magnetic field strength was set to be 16.5 mT to prevent micro-disks from stepping out. This batch of micro-disks was produced in the summer, and its magnetic moment is not as high as those produced in the winter, so a higher-than-usual field was used. The magnification of the zoom lens was 0.57x.

The experiments for phase transitions were performed for a collective of 251 spinning micro-disks of 6-fold symmetry and for 267 and 198 micro-disks of 5-fold and 4-fold symmetries, respectively (**Fig. 4**). Videos were recorded for 15 minutes continuously. The magnification of the zoom lens was 0.57x. The rotation speed and field strength were set according to the list described below for micro-disks of 6-fold symmetry. Field strength values in parentheses are for micro-disks of 4-fold and 5-fold symmetries.

1. Ω = 30 rps – 20 rps in steps of 5 rps, B = 10 mT (14 mT), 10 s
2. Ω = 18 rps – 10 rps in steps of 2 rps, B = 10 mT (14 mT), 10 s
3. Ω = 9 rps – 1 rps in steps of 1 rps, B = 10 mT (14 mT), 10 s

4. Ω = 1 rps, B = 1 mT (10 mT), 30 s
5. Ω = 0.75 rps, B = 1 mT (10 mT), 30 s
6. Ω = 0.5 rps, B = 1 mT (5 mT), 60 s
7. Ω = 0.25 rps, B = 1 mT (1 mT), 60 s
8. Mix 1 [Ω = 1 rps and B = 3 mT (3 mT) for 1 s] and [Ω = 0.25 rps and B = 0.5 mT (0.5 mT) for 4 s], 90 s
9. Mix 2 [Ω = 5 rps and B = 3 mT (3 mT) for 1 s] and [Ω = 0.25 rps and B = 0.5 mT (0.5 mT) for 4 s], 400 s
10. Ω = 0.25 rps, B = 1 mT (1 mT), 60 s

**Calculation of order parameters**

The hexatic order parameter was calculated according to

$$\psi_6 = \frac{\sum_k \exp(i6\vartheta_k)}{K}, \tag{4}$$

where $K$ is the number of one micro-disk's neighbours; $k$ is the neighbour index; $\vartheta_k$ is the polar angle of the vector from the micro-disk to its neighbour $k$.

**Calculation of radial distribution functions**

We consider each micro-disk in turn and count the number of micro-disks within a circular band of width $R$. Then we sum the counts of all the micro-disks and divide the total count by the total area of the circular band and by the total number of micro-disks. We repeat this process from radial distance of $2R$ to $100R$.

**Model for pairwise interactions**

If the edge-edge distance $d \geq$ lubrication threshold (=15 μm, or 0.1R)

$$\frac{d\boldsymbol{r_i}}{dt} = \sum_{j\neq i}(6\pi\mu R)^{-1}\left(F_{mag-on,\,i,j}\big(r_{ji},\varphi_{ji}\big) + F_{cap,\,i,j}\big(r_{ji},\varphi_{ji}\,\big) + \frac{\rho\omega^2R^7}{r_{ji}^3}\right)\hat{r}_{ji} + \sum_{j\neq i}\left(\frac{F_{mag-off,\,i,j}\big(r_{ji},\varphi_{ji}\big)}{6\pi\mu R} - \frac{R^3\omega}{r_{ji}^2}\right)\hat{r}_{ji} \times \hat{z},\ i = 1,2 \tag{5}$$

$$\frac{d\alpha_i}{dt} = \frac{mB\sin(\theta-\alpha_i)}{8\pi\mu R^3} + \sum_{j\neq i}\frac{T_{mag-d,i,j}\big(r_{ji},\ \varphi_{ji}\big) + T_{cap,i,j}\big(r_{ji},\ \varphi_{ji}\big)}{8\pi\mu R^3},\ \ i = 1,2 \tag{6}$$

where $\boldsymbol{r_i}$ and $\boldsymbol{r_j}$ are the position vectors of micro-disks;

$\boldsymbol{r_{ji}} = \boldsymbol{r_i} - \boldsymbol{r_j}$ is the vector pointing from the centre of micro-disk *j* to the centre of micro-disk *i*;

$\alpha_i$ and $\alpha_j$ are the orientations of micro-disks;

*d* is the edge-edge distance;

$\varphi_{ji}$ is the angle of dipole moment with respect to $\boldsymbol{r_{ji}}$. It is assumed to be the same for both micro-disks, as $\varphi_{ji} = \varphi_i = \varphi_j$ in scheme S1 in the section on magnetic dipole force and torque calculation.

$\omega$ is the instantaneous spin speed of micro-disks;

$\theta = \Omega t$ is the orientation of the magnetic field;

$\Omega$ is the rotation speed of the magnetic field;

$R$ is the radius of micro-disk (150 μm);

$\mu$ is the dynamic viscosity of water ($10^{-3}$ Pa·s);

$\rho$ is the density of water ($10^3$ kg/m$^3$);

$m$ is the magnetic dipole moment of the micro-disks ($10^{-8}$ A·m$^2$);

$B$ is the magnetic field strength (10 mT);

$F_{mag-on,\,i,j}$ and $F_{mag-off,\,i,j}$ are the magnetic dipole force on and off the centre-to-centre axis, respectively, and they are functions of $r_{ji}$ and $\varphi_{ji}$; (See the section on magnetic dipole force and torque calculation for details)

$T_{mag-d,i,j}$ is the magnetic dipole torque, and it is a function of $r_{ji}$ and $\varphi_{ji}$; (See the section on magnetic dipole force and torque calculation for details)

$F_{cap,\,i,j}$ is the capillary force, and it is a function of $r_{ji}$ and $\varphi_{ji}$ and embeds the symmetry of a micro-disk: (See section on capillary force and torque calculation for details)

$T_{cap,i,j}$ is the capillary torque, and it is a function of $r_{ji}$ and $\varphi_{ji}$ and embeds the symmetry of a micro-disk. (See section on capillary force and torque calculation for details)

If the edge-edge distance $d$ < lubrication threshold (=15 μm, or 0.1R) and d ≥ 0,

$$\mu \frac{d\boldsymbol{r_i}}{dt} = \sum_{j\neq i} A\left(\frac{d}{R}\right)\Bigg(F_{mag-on,\,i,j}\big(r_{ji},\varphi_{ji}\big) + F_{cap,\,i,j}\big(r_{ji},\varphi_{ji}\,\big) + \frac{\rho\omega^2 R^7}{r_{ji}^3}\Bigg)\hat{r}_{ji} + \sum_{j\neq i} B\left(\frac{d}{R}\right) F_{mag-off,\,i,j}\big(r_{ji},\varphi_{ji}\big)\,\hat{r}_{ji}\times\hat{\mathrm{z}} + \sum_{j\neq i} C\left(\frac{d}{R}\right) mB\sin(\theta-\alpha_i)\,\hat{r}_{ji}\times\hat{\mathrm{z}},\ i=1,2 \tag{7}$$

$$\mu\frac{d\alpha_i}{dt} = G\left(\frac{d}{R}\right) mB\sin(\theta-\alpha_i) + \sum_{j\neq i} G\left(\frac{d}{R}\right) T_{mag-d,i,j}\big(r_{ji},\ \varphi_{ji}\big) + T_{cap,i,j}\big(r_{ji},\ \varphi_{ji}\big)\,,\ \ i=1,2 \tag{8}$$

where the coefficients are defined as the following (S. Kim, S. J. Karrila, *Microhydrodynamics* (Butterworth-Heinemann, 1991)):

$$A(x) = \frac{x(-0.285524x + 0.095493x\ln(x) + 0.106103\,)}{R} \tag{9}$$

$$B(x) = \frac{\left(0.0212764\ln\left(\frac{1}{x}\right) + 0.157378\right)\ln\left(\frac{1}{x}\right) + 0.269886}{R\left(\ln\left(\frac{1}{x}\right)\left(\ln\left(\frac{1}{x}\right) + 6.0425\right) + 6.32549\right)} \tag{10}$$

$$C(x) = \frac{\left(-0.0212758\ln\left(\frac{1}{x}\right) - 0.089656\right)\ln\left(\frac{1}{x}\right) + 0.0480911}{R^2\left(\ln\left(\frac{1}{x}\right)\left(\ln\left(\frac{1}{x}\right) + 6.0425\right) + 6.32549\right)} \tag{11}$$

$$G(x) = \frac{\left(0.0212758\ln\left(\frac{1}{x}\right) + 0.181089\right)\ln\left(\frac{1}{x}\right) + 0.381213}{R^3\left(\ln\left(\frac{1}{x}\right)\left(\ln\left(\frac{1}{x}\right) + 6.0425\right) + 6.32549\right)} \tag{12}$$

**Model for many-disk interactions**

If the edge-edge distance $d_{ji} \geq$ lubrication threshold (=15 μm, or 0.1R),

$$\frac{d\boldsymbol{r_i}}{dt} = \sum_{j\neq i}(6\pi\mu R)^{-1}\left(F_{mag-on,\,i,j}\left(r_{ji},\varphi_{ji}\right) + F_{cap,\,i,j}\left(r_{ji},\varphi_{ji}\ \right) + \frac{\rho\omega^2 R^7}{r_{ji}^3}\right)\hat{r}_{ji} + \sum_{j\neq i}\left(\frac{F_{mag-off,\,i,j}\left(r_{ji},\varphi_{ji}\right)}{6\pi\mu R} - \frac{R^3\omega}{r_{ji}^2}\right)\hat{r}_{ji}\times\hat{\mathrm{z}} + \frac{\rho\omega_i^2 R^7}{6\pi\mu R}\cdot\left(\left(\frac{1}{d_{toLeft}^3} - \frac{1}{d_{toRight}^3}\right)\hat{x} + \left(\frac{1}{d_{toBottom}^3} - \frac{1}{d_{toTop}^3}\right)\hat{y}\right),\ i = 1,2,\dots \tag{13}$$

$$\frac{d\alpha_i}{dt} = \frac{mB\sin(\theta-\alpha_i)}{8\pi\mu R^3} + \sum_{j\neq i}\frac{T_{mag-d,i,j}\left(r_{ji},\ \varphi_{ji}\right) + T_{cap,i,j}\left(r_{ji},\ \varphi_{ji}\right)}{8\pi\mu R^3},\ \ i = 1,2,\dots \tag{14}$$

where $\boldsymbol{r}_{center}$ is the position vector of the centre of the arena;

$R_{arena}$ is the radius of the arena;

$d_{toLeft}, d_{toRight}, d_{toBottom}$, and $d_{toTop}$ are the distances of a micro-disk to the four sides of the arena.

If the edge-edge distance $d_{ji}$ < lubrication threshold (=15 μm, or 0.1R) and $d_{ji} \geq 0$,

$$
\begin{aligned}
\mu \frac{d\boldsymbol{r_i}}{dt} = \sum_{j\neq i} A\left(\frac{d_{ji}}{R}\right)\Bigg( & F_{mag-on,\,i,j}\big(r_{ji},\varphi_{ji}\big) \\
& + F_{cap,\,i,j}\big(r_{ji},\varphi_{ji}\,\big) + \frac{\rho\omega^2 R^7}{r_{ji}^3}\Bigg)\hat{r}_{ji} \\
& + \sum_{j\neq i} B\left(\frac{d_{ji}}{R}\right) F_{mag-off,\,i,j}\big(r_{ji},\varphi_{ji}\big)\,\hat{r}_{ji}\times\hat{\mathrm{z}} \\
& + \sum_{j\neq i} C\left(\frac{d_{ji}}{R}\right) mB\sin(\theta-\alpha_i)\,\hat{r}_{ji}\times\hat{\mathrm{z}} \\
& + \frac{\rho\omega_i^2 R^7}{6\pi R}\Bigg(\left(\frac{1}{d_{toLeft}^3}-\frac{1}{d_{toRight}^3}\right)\hat{x} \\
& + \left(\frac{1}{d_{toBottom}^3}-\frac{1}{d_{toTop}^3}\right)\hat{y}\Bigg),\ i=1,2,\ldots
\end{aligned} \tag{15}
$$

$$
\begin{aligned}
\mu \frac{d\alpha_i}{dt} = G\left(\frac{d_{smallest}}{R}\right) & mB\sin(\theta-\alpha_i) \\
& + \sum_{j\neq i} G\left(\frac{d_{ji}}{R}\right) T_{mag-d,i,j}\big(r_{ji},\ \varphi_{ji}\big) \\
& + T_{cap,i,j}\big(r_{ji},\ \varphi_{ji}\big)\,,\ \ i=1,2,\ldots
\end{aligned} \tag{16}
$$

If the edge-edge distance $d_{ji} < 0$, a repulsion term is added to the force equation,

$$
\begin{aligned}
\mu \frac{d\boldsymbol{r_i}}{dt} = \sum_{j\neq i} A(\varepsilon)\Bigg( & F_{mag-on,\,i,j}\big(2R,\varphi_{ji}\big) + F_{cap,\,i,j}\big(2R,\varphi_{ji}\,\big) \\
& + \frac{\rho\omega^2 R^7}{r_{ji}^3}\Bigg)\hat{r}_{ji} + \sum_{j\neq i}\frac{F_{wallRepulsion}}{6\pi R}\frac{-d_{ji}}{R}\hat{r}_{ji} \\
& + \sum_{j\neq i} B(\varepsilon) F_{mag-off,\,i,j}\big(2R,\varphi_{ji}\big)\,\hat{r}_{ji}\times\hat{\mathrm{z}} \\
& + \sum_{j\neq i} C(\varepsilon)\, mB\sin(\theta-\alpha_i)\,\hat{r}_{ji}\times\hat{\mathrm{z}} \\
& + \frac{\rho\omega_i^2 R^7}{6\pi R}\Bigg(\left(\frac{1}{d_{toLeft}^3}-\frac{1}{d_{toRight}^3}\right)\hat{x} \\
& + \left(\frac{1}{d_{toBottom}^3}-\frac{1}{d_{toTop}^3}\right)\hat{y}\Bigg),\ i=1,2,\ldots
\end{aligned} \tag{17}
$$

$$\mu \frac{d\alpha_i}{dt} = G(\varepsilon) m B \sin(\theta - \alpha_i) + \sum_{j \neq i} G(\varepsilon) T_{mag-d,i,j}\left(2R,\ \varphi_{ji}\right) + T_{cap,i,j}\left(2R,\ \varphi_{ji}\right),\ \ i = 1,2,\dots \tag{18}$$

where $\varepsilon$ is a small number ($10^{-10}$ μm/$R$); $F_{wallRepulsion}$ is set to be $10^{-7}$ N.

**Magnetic dipole force and torque calculation**

The geometry of interaction between two magnetic dipoles is shown in scheme 1 below.

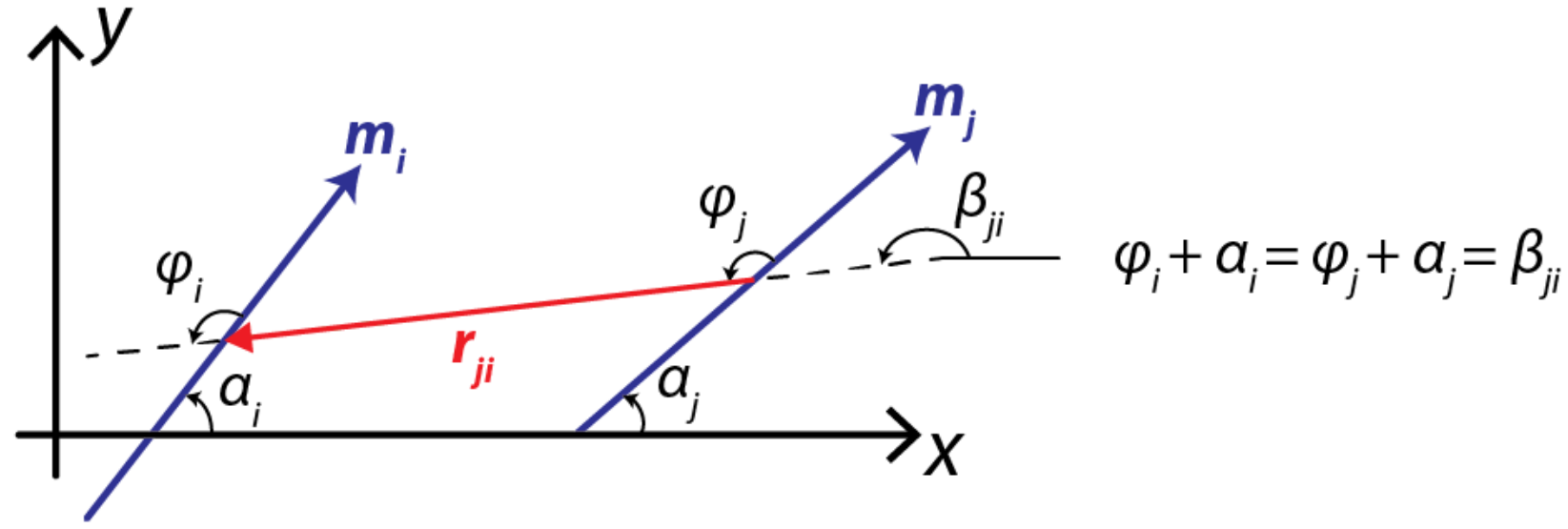

Scheme S1. The geometry of dipole-dipole interaction. The blue arrows are the directions of the magnetic dipole of micro-disks. The red arrow is the centre-to-centre axis of micro-disks.

The force by dipole *j* on dipole *i* (K. W. Yung, P. B. Landecker, D. D. Villani, An Analytic Solution for the Force Between Two Magnetic Dipoles. *Magn. Electr.* Sep. 9, 39–52 (1998)):

$$\boldsymbol{F}_{by\ j\ on\ i} = \frac{3\mu_0 m_j m_i}{4\pi r_{ji}^4}\Big(\hat{r}_{ji}(\widehat{m}_j \cdot \widehat{m}_i) + \widehat{m}_j(\hat{r}_{ji} \cdot \widehat{m}_i) + \widehat{m}_i(\hat{r}_{ji} \cdot \widehat{m}_j) - 5\hat{r}_{ji}(\hat{r}_{ji} \cdot \widehat{m}_j)(\hat{r}_{ji} \cdot \widehat{m}_i)\Big), \tag{19}$$

$$\boldsymbol{F}_{by\ j\ on\ i} = \frac{3\mu_0 m_j m_i}{4\pi r_{ji}^4}\left(\cos(\varphi_j - \varphi_i)\,\hat{r}_{ji} + \cos\,(\varphi_i)\widehat{m}_j + \cos\,(\varphi_j)\widehat{m}_i - 5\cos(\varphi_i)\cos(\varphi_j)\,\hat{r}_{ji}\right), \tag{20}$$

where

the hat $\frown$ denotes a unitized vector;

$\boldsymbol{r}_{ji} = \boldsymbol{r}_i - \boldsymbol{r}_j$ is the vector pointing from disk *j* to disk *i*;

$\mu_0 = 4\pi \times 10^{-7} N/A^2$ is the vacuum permeability;

$m_i$ and $m_j$ are the magnetic moments of micro-disks.

$\alpha_i$ and $\alpha_j$ are defined in Scheme 1.

With the geometric relations

$$\hat{m}_i = \cos(\varphi_i)\,\hat{r}_{ji} + \sin(\varphi_i)\,\hat{r}_{ji} \times \hat{z}\,, \tag{21}$$

$$\hat{m}_j = \cos(\varphi_j)\,\hat{r}_{ji} + \sin(\varphi_j)\,\hat{r}_{ji} \times \hat{z}\,, \tag{22}$$

the force equation becomes

$$\begin{aligned} \boldsymbol{F}_{by\ j\ on\ i} = \frac{3\mu_0 m_j m_i}{4\pi r_{ji}^4}\Big(&\big(-2\cos(\varphi_i)\cos(\varphi_j) \\ &+ \sin(\varphi_i)\sin(\varphi_j)\big)\hat{r}_{ji} \\ &+ \big(\cos(\varphi_i)\sin(\varphi_j) + \cos(\varphi_j)\sin(\varphi_i)\big)\hat{r}_{ji} \\ &\times \hat{z}\Big)\,. \end{aligned} \tag{23}$$

Set $\varphi_i = \varphi_j = \varphi_{ji}$, then

$$\begin{aligned} \boldsymbol{F}_{by\ j\ on\ i} = \frac{3\mu_0 m_j m_i}{4\pi r_{ji}^4}\Big(&\big(1 - 3\cos^2(\varphi_{ji})\big)\hat{r}_{ji} \\ &+ 2\cos(\varphi_{ji})\sin(\varphi_{ji})\,\hat{r}_{ji} \times \hat{z}\Big)\,, \end{aligned} \tag{24}$$

$$F_{mag-on,\,i,j}(r_{ji}, \varphi_{ji}) = \frac{3\mu_0 m_j m_i}{4\pi r_{ji}^4}\big(1 - 3\cos^2(\varphi_{ji})\big)\,, \tag{25}$$

$$F_{mag-off,\,i,j}(r_{ji}, \varphi_{ji}) = \frac{3\mu_0 m_j m_i}{4\pi r_{ji}^4}\big(2\cos(\varphi_{ji})\sin(\varphi_{ji})\big)\,. \tag{26}$$

The torque by dipole *j* on dipole *i* (P. B. Landecker, D. D. Villani, K. W. Yung, Analytic solution for the torque between two magnetic dipoles. *Magn. Electr. Sep.* 10, 29–33 (1999)):

$$\boldsymbol{T}_{by\ j\ on\ i} = \frac{\mu_0 m_j m_i}{4\pi r_{ji}^3}\left[3(\hat{m}_j \cdot \hat{r}_{ji})(\hat{m}_i \times \hat{r}_{ji}) + (\hat{m}_j \times \hat{m}_i)\right]\,, \tag{27}$$

$$\begin{aligned} \boldsymbol{T}_{by\ j\ on\ i} = \frac{\mu_0 m_j m_i}{4\pi r_{ji}^3}\big[&3\cos(\varphi_j)\sin(\varphi_i)\,\hat{z} \\ &+ \sin(\varphi_i - \varphi_j)\hat{z}\big]\,. \end{aligned} \tag{28}$$

Set $\varphi_i = \varphi_j = \varphi_{ji}$, then

$$\boldsymbol{T}_{by\ j\ on\ i} = \frac{\mu_0 m_j m_i}{4\pi r_{ji}^3}\big(3\cos(\varphi_{ji})\sin(\varphi_{ji})\,\hat{z}\big)\,. \tag{29}$$

**Capillary force and torque calculation**

The area of the air-water interface with two static micro-disks can be calculated analytically, and hence the surface energy is just the area times the surface tension of water. The surface energy is a function of the separation distance and the orientations of two micro-disks. The capillary force and torque are calculated from the derivatives of this energy with respect to the separation distance and the orientation angle of the micro-disks, respectively.

In general, any edge undulation profile $H(\vartheta)$ can be expressed as the sum of its Fourier modes:

$$H(\theta) = \sum_{n=0}^{\infty} A_n \sin(n\vartheta) \,, \tag{30}$$

where $A_n$ are the Fourier coefficients, and $\vartheta$ is the polar angle.

For two micro-disks, the surface energy is the summation of all modes of both micro-disks. Each mode can be calculated exactly in bipolar coordinates (K. D. Danov, P. A. Kralchevsky, B. N. Naydenov, G. Brenn, Interactions between particles with an undulated contact line at a fluid interface: Capillary multipoles of arbitrary order. *J. Colloid Interface Sci.* 287, 121 (2005)):

$$\frac{E_{m1,m2}}{\pi\sigma} = H_1^2 S_{m1} + H_2^2 S_{m2} - H_1 H_2 G_{m1,m2} \cos(m_1\varphi_1 + m_2\varphi_2) \,, \tag{31}$$

where σ is the surface tension of water;

$H_i$ is the amplitude of the sinusoid on micro-disk $i$, and $i = 1, 2$ is the index of the micro-disk;

$\varphi_i$ is the orientation of the micro-disk $i$, as in Scheme 1.

$m_i$ is the mode of the micro-disk $i$;

$S_n$ and $G_{n,m}$ are given below:

$$S_n = \sum_{k=1}^{\infty} \frac{k}{2} \coth\left(2k \operatorname{acosh}\left(\frac{d}{2R} + 1\right)\right) \Xi^2\left(k, n, \operatorname{acosh}\left(\frac{d}{2R} + 1\right)\right) \,, \tag{32}$$

$$G_{n,m} = \sum_{k=1}^{\infty} \frac{k\Xi\left(k, n, \operatorname{acosh}\left(\frac{d}{2R} + 1\right)\right) \Xi\left(k, m, \operatorname{acosh}\left(\frac{d}{2R} + 1\right)\right)}{\sinh\left(2k \operatorname{acosh}\left(\frac{d}{2R} + 1\right)\right)} \,, \tag{33}$$

$$\Xi(n, m, v) = m \sum_{k=0}^{\min(m,n)} \frac{(-1)^{m-k}(m + n - k - 1)!}{(m - k)!\,(n - k)!\,k!} e^{-(m+n-2k)v} \,, \tag{34}$$

where acosh() is the inverse of the hyperbolic cosine function;

$R$ is the radius of the micro-disk;

$d$ is the edge-to-edge distance.

If $\alpha_1 = \alpha_2 = \alpha$, the total energy then is

$$E(d,\alpha) = \sum_{m1,m2} E_{m1,m2}(d,\alpha)\,. \tag{35}$$

The capillary force and torques then are calculated as

$$F(d,\alpha) = -\frac{\partial E}{\partial d}, \tag{36}$$

$$T(d,\alpha) = -\frac{\partial E}{\partial \alpha}. \tag{37}$$

**Supplementary notes on the scaling relations**

1. *Magnetic dipole-dipole interactions*:

The angle-averaged magnetic force was calculated by averaging equation 25 with respected to $\phi_{ji}$ as:

$$F_{mag}(r) = -\frac{3\mu_0 m^2 \pi^2}{4d^4} = -\frac{3}{4}\pi^2 \mu_0 \rho_m^2 \times \left(\frac{1}{d/R}\right)^4 \tag{38}$$

where $m = \rho_m \pi R^2$ is the total magnetic moment of each micro-disk;

$\rho_m$ is the magnetic moment per unit area of the disk and is estimated to be ~0.1A from **Fig. S2G**, and we chose the thickness of 500 nm so that the sputtering process could be conveniently finished within one day;

$R$ is the radius of the disk;

$d$ is the centre-to-centre distance between micro-disks.

Because the magnetic moment is due to the cobalt thin film and scales with its surface area ($m \propto R^2$), the magnitude of interaction depends only on the factor of $d/R$.

2. *Hydrodynamic interactions*:

The hydrodynamic lift force is calculated as:

$$F_{hydro}(r) = \frac{\rho\omega^2 R^7}{d^3} = \frac{\rho(\omega R^2)^2}{(d/R)^3} = \rho\nu^2 \frac{\mathrm{Re}^2}{(d/R)^3} \tag{39}$$

where $\rho$ is the density of water;

$\omega$ is the spin speed in rad/s;

$\nu$ is the kinematic viscosity of water;

Re is the Reynolds number.

3. *Capillary interactions*:

The numerical values for the capillary forces were calculated using equation 36 (see the section on Capillary force and torque calculation). The angle-averaged value of capillary forces were then calculated as

$$F_{cap}(d) = \frac{1}{2\pi} \cdot \sum_{\alpha=0}^{2\pi} [F_{cap}(d,\alpha) \cdot \Delta\alpha], \tag{40}$$

where $\Delta\alpha = 1^o$ is the step size.

Micro-disks are assumed to have six cosinusoidal profiles of amplitude 2 µm and an arc angle of 30 degrees along the circumference. The numerical relations for the capillary forces at fixed *d*/*R* are as follows:

$F_{cap}(d = 2.25R) = 10^{-12.4}\mathrm{N} \cdot \mathrm{m} \cdot R^{-1.0}$, $[F_{cap}] = N$, $[R] = m$

$F_{cap}(d = 2.5R) = 10^{-13.5}\mathrm{N} \cdot \mathrm{m} \cdot R^{-1.0}$

$F_{cap}(d = 2.75R) = 10^{-14.4}\mathrm{N} \cdot \mathrm{m} \cdot R^{-1.0}$

$F_{cap}(d = 3R) = 10^{-15.2}\mathrm{N} \cdot \mathrm{m} \cdot R^{-1.0}$

From these relations, one can deduce a relation

$$F_{cap}(d) = 10^{-4.1-3.7\left(\frac{d}{R}\right)} \cdot R^{-1.0} \tag{41}$$

where *d* is the centre-centre distance between disks.

## Supplementary notes on the Hamiltonian approach

The one–dimensional (1D) Hamiltonian of pairwise interaction in the pattern of many disks is calculated as:

$$H(d) = E_{mag-dp}(d) + E_{cap}(d) + E_{hydro}(d) + \Gamma \cdot E_{bd}(d) \tag{42}$$

where $E_{cap}(d)$ is the angle-averaged capillary energy;

$E_{mag-dp}(d)$ is the angle-averaged magnetic dipole-dipole energy;

$E_{hydro}(l) = -\int F_{hydro}(l) \cdot dl \ = \frac{\rho R^7 \omega^2}{2l^2}$ is the effective hydrodynamic energy;

$d$ is the centre-centre distance between micro-disks;

Γ is a fitting parameter, and its value is 10 for all the spin speeds.

Derivation of the mean field energy term

The geometry of interaction between two micro-disks is shown in Scheme 2 below.

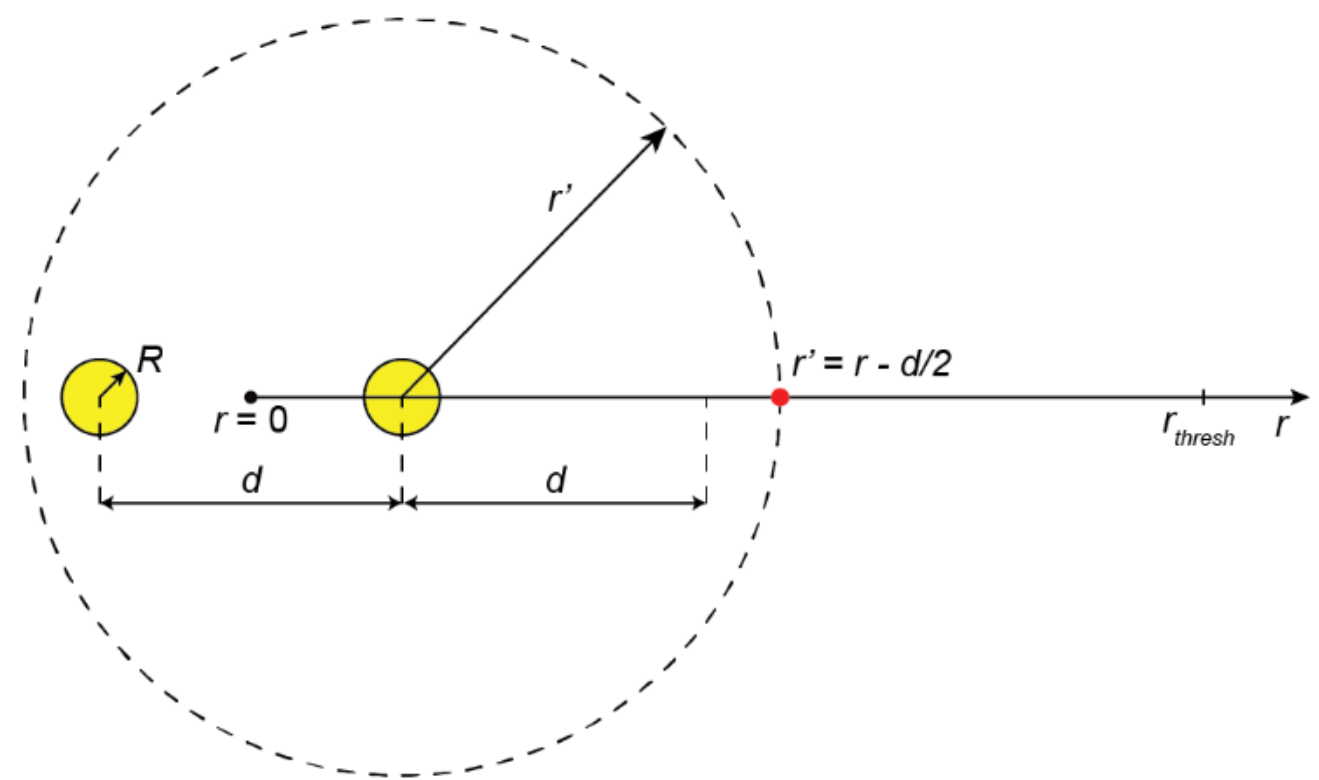

Scheme S2: The geometry used for the construction of a mean field energy term to account for the interactions between one pair of micro-disks with all the rest micro-disks. When considering the micro-disk on the right, the area density of micro-disks is assumed to be uniform from $r = 3d/2$ to $r = r_{thresh}$. $r_{thresh}$ is half of the edge length of the physical boundary.

The mean field term is calculated according to Scheme S2. All the micro-disks (except for the pair under consideration) are assumed to be uniformly distributed in $r \in [\frac{3d}{2}, r_{thresh}]$, where *r* is the distance from the centre of mass of the pair of micro-disks. In Scheme S2, we consider an imaginary circle of micro-disks (the dotted circle). This circle is centred on the right micro-disk, and its radius is $r'$. According to our assumption, for $d \le r' \le r_{thresh} - \frac{d}{2}$, the micro-disks are uniformly distributed, and there are no micro-disks beyond $r_{thresh}$. The interaction energies between the right (yellow) micro-disk with all the micro-disks whose centres are on the imaginary circle are the same (assuming angle-averaged interactions). Then we map all the points on the circle to one point on *r*-axis, $r' = r - d/2$, the red dot in Scheme 2. We sum all the interaction energies from $r' = d$ to $r' = r_{thresh} - \frac{d}{2}$, and consider this sum to be the mean field energy acting the yellow micro-disk on the right. Because energy is additive, we multiply this value by 2 and consider the product to be the effective boundary energy of the pair of micro-disks separated by a distance *d*.

The probability distribution as a function of *r* and *d* is given as:

$$P(r,d) = \frac{2\pi\left(r - \frac{d}{2}\right)}{\sum_{r=\frac{3d}{2}}^{r_{thresh}} 2\pi\left(r - \frac{d}{2}\right) \cdot \Delta r} \tag{43}$$

It is normalized as,

$$\sum_{r=\frac{3d}{2}}^{r_{thresh}} P(r) \cdot \Delta r = 1\,, \tag{44}$$

where $N$ is the number of disks in the system and $\Delta r$ is the step size.

The mean field energy $E_{mf}(d)$ is calculated as the mean of all interaction energies between the pair of micro-disks under consideration and all other micro-disks in the arena:

$$E_{mf}(d) = \sum_{r=\frac{3d}{2}}^{r_{thresh}} 2 \cdot \left(\left[E_{cap}\left(r - \frac{d}{2}\right) + E_{mag-dp}\left(r - \frac{d}{2}\right) + E_{hydro}\left(r - \frac{d}{2}\right)\right] \cdot P(r) \cdot \Delta r\right) \quad (45)$$

where the summation is performed in steps of 1 μm.

The algorithm we use for fitting $\beta$ is as follows:

1. cost = KLD (Kulback-Leibler divergence, see below for the definition). For each rotation speed Ω in [12, 70] rps with a step size of 1 rps
2. Calculate the terms in the Hamiltonian (different energy contributions)
3. For $\beta$ in [1, 10000) * $10^{11}$ with a step size of 1*$10^{11}$ (unit of $\beta$ is $J^{-1}$)
    1) Calculate the complete Hamiltonian using eq (42)
    2) Calculate the histogram of the probability distribution $p(\, d;\ \beta) = e^{-\beta H(d;\, r_{thresh})}$
    3) Calculate KL divergence (KLD) from the experimental neighbour distance distribution
    4) If KLD is less than the previous minimum value, save the current KLD as the new KLD minimum
4. Return $\beta$ and $p(\, d;\ \beta)$ corresponding to the minimum value of KLD

Implementation

We calculate the value of Hamiltonian for $d \in [300, 1500)\ \mu m$ with a step size of 1 $\mu m$. The reason for using this range of $d$ is because, for the experimental neighbour distance distribution, we do the binning till $d = 10R$.

Capillary interaction calculations

The capillary interaction energy was calculated as mentioned in the section on Capillary force and torque calculation and angle-averaged values were calculated similar to equation 40.

Magnetic dipole-dipole interaction calculations

Magnetic dipole-dipole interaction energy was calculated by integrating the force in equation 38 with respect to $r$.

Hydrodynamic lift force interaction calculations

Hydrodynamic lift force interaction energy was calculated by integrating the force in equation 39 with respect to $r$. This was done for each $\omega \in [12, 70]$ rps.

Boundary energy calculations

For boundary energy calculations, equation 45 was used for each $\Omega \in [12, 70]\ rps$.

Function for calculating the histogram

A routine for calculating the histogram was written. It takes in the 1D array containing the value of $p(d;\ \beta) = e^{-\beta \cdot H(d;\ r_{thresh})}$ and returns the histogram of $d$. The bin size and bin edges used are the same as those used for the experimental neighbour distance distribution. Bin edges were [2R, 10R) with a bin size of 0.5R and the last bin was from 10R to 100R.

Kulback-Leibler Divergence calculation

Kulback-Leibler Divergence is defined as

$$D_{KL}(P||Q) = \Sigma P(x) \ln\left(\frac{P(x)}{Q(x)}\right) \tag{46}$$

where P(x) and Q(x) are two distributions

It was calculated using the entropy function of the scipy library in python. Histogram of neighbour distance from experiments was used as $P$(x) and the histogram calculated from $p(d;\ \beta) = e^{-\beta \cdot H(d;\ r_{thresh})}$ was used as $Q$(x). A small value of $10^{-3}$ was added to the Q(x) to avoid division by zero problem.

**Supplementary note on the Monte Carlo simulations**

The algorithm used is as follows:

- Start with all $N$ disks in randomly generated positions ensuring no overlap.
- for each time step:
    - for each disk
        1. Generate a random movement vector within the arena
        2. Check if the new position of the disk overlaps with any other disks
        3. Calculate KLD for the distribution(s)
        4. Check if after the movement, the distributions converge
        5. If the KLD decreases: accept the move
        6. Else: generate a random number, $p$ in [0,1)
            1. If $p < e^{-\beta_{MC} KLD_{diff}}$ : accept the move
            2. Else reject the move

$\beta_{MC}$ was chosen as the inverse of standard deviations of KLD between experimental frames and was same for all four representative patterns.

$KLD_{diff}$ is the KLD difference from experiments corresponding to the distribution of neighbor distances.

The simulations were run using only neighbour distances for each of the four spin speeds. The algorithms were for ~150k steps. Simulations were repeated at least seven times for each of the four spin speeds.

**Acknowledgments**

We thank Z. Davidson, U. Culha, X. Dong, A. Karacakol, and P.M. Bonetti for discussions, X. Hu for magnetic calculations, G. Ricther and F. Thiele for help on sputtering and 3D printing of arenas, J. Fiene and B. Wright for 3D printing of the coil frames, N. K.-Subbaiah for help on Nanoscribe 3D micro-printing, and M.S. Santhanam for feedback on the manuscript. WW and ZW thank the high-performance computing center at SJTU for support on simulations. **Funding:** We thank the Max Planck Society for funding; WW and HG thank the Alexander von Humboldt Foundation for fellowships; GG thanks the International Max Planck Research School for Intelligent Systems (IMPRS-IS) for support. W.W. thanks UM-SJTU JI for support. **Author contributions:** WW, GG, PM, VK, and MS designed the study; WW, GG, and PH performed the experiments; DS, HG, and WW designed the experimental setup; LK, EL, and WW constructed the pairwise numerical models; GG, PM, WW and VK constructed the Hamiltonian model; WW, GG, ZW and VK performed the Monte Carlo simulations; WW, GG, PM, VK, and PH analyzed the data; WW and GG wrote the manuscript; all authors contributed to the editing of the manuscript; CH provided constructive feedback on the work and advice; MS supervised the research. **Competing interests:** Authors declare no competing interests. **Data and materials availability:** All data are available in the main text or supplementary materials. Video processing and simulation codes will be available in open source in GitHub.

## Figures and Tables

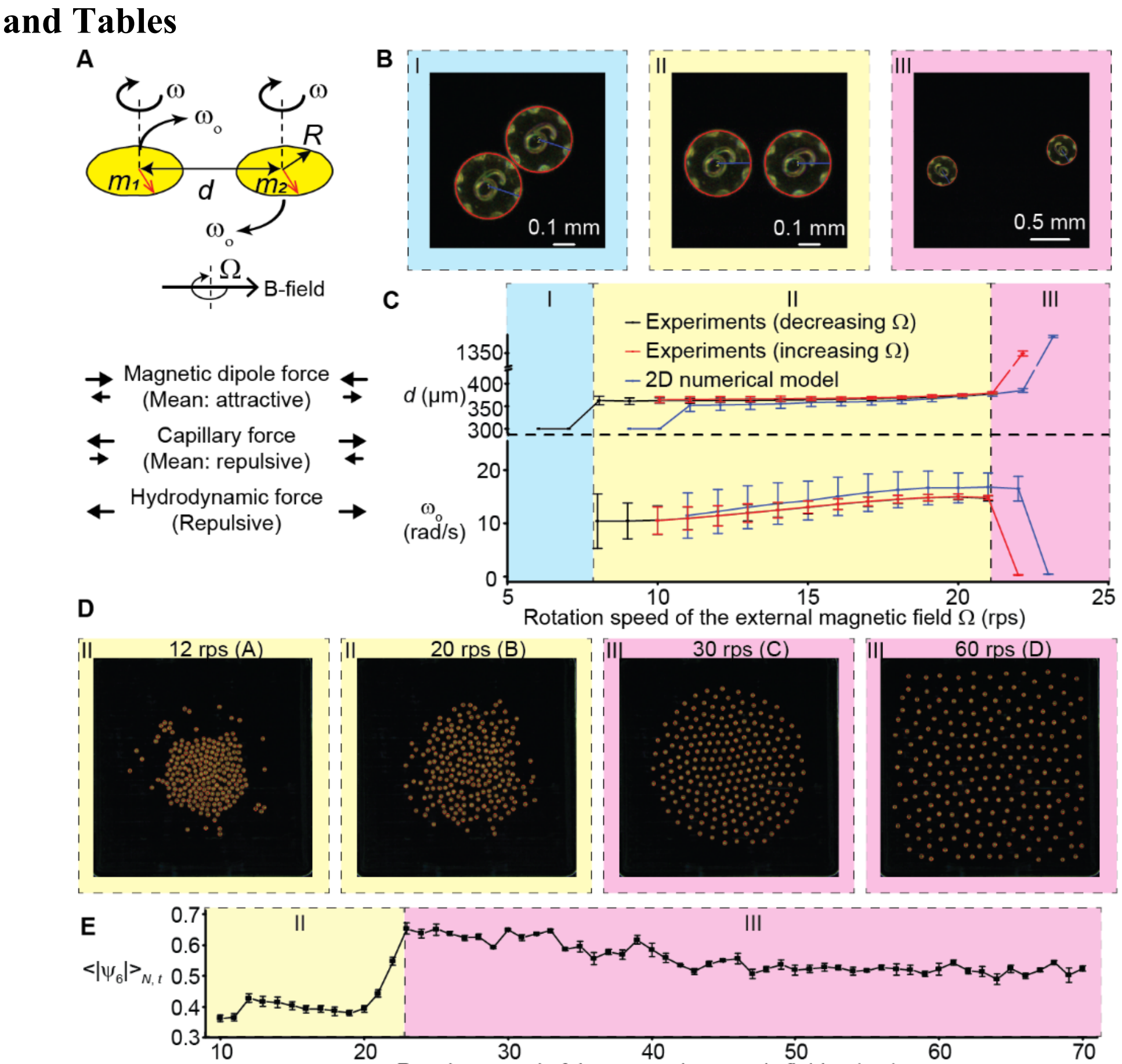

**Fig. 1 The tunable balance of local pairwise interactions produces global patterns with varying degree of orders.** (**A**) The scheme of pairwise interactions shows the three pairwise forces: the magnetic dipole-dipole force (whose average after one full rotation is attractive), the capillary force (whose average after one full rotation is repulsive), and the hydrodynamic lift force (always repulsive). (**B**) Three representative images of a pair of micro-disks showing the attached state (I), the orbiting state (II), and the decoupled state (III). (**C**) The center-center distance $d$ and the orbiting speed $\omega_o$ as functions of the rotation speed of the magnetic field **Ω** (in the unit of revolutions per second, rps). The experimental curves are labelled with the direction of change in **Ω**. The black (red) curve shows the assembling (decoupling) transition. The blue curves are based on the 2D numerical model of the equations of motion. (**D**) Experimental images of 218 micro-disks showing representative patterns. The background colours indicate regions II – III and show the correspondences between pairwise states and patterns of many micro-disks. (**E**) The averaged norm of the hexatic order parameters $<|\psi_6|>_{N,t}$ as a function of the rotation speed of the external magnetic field **Ω**.

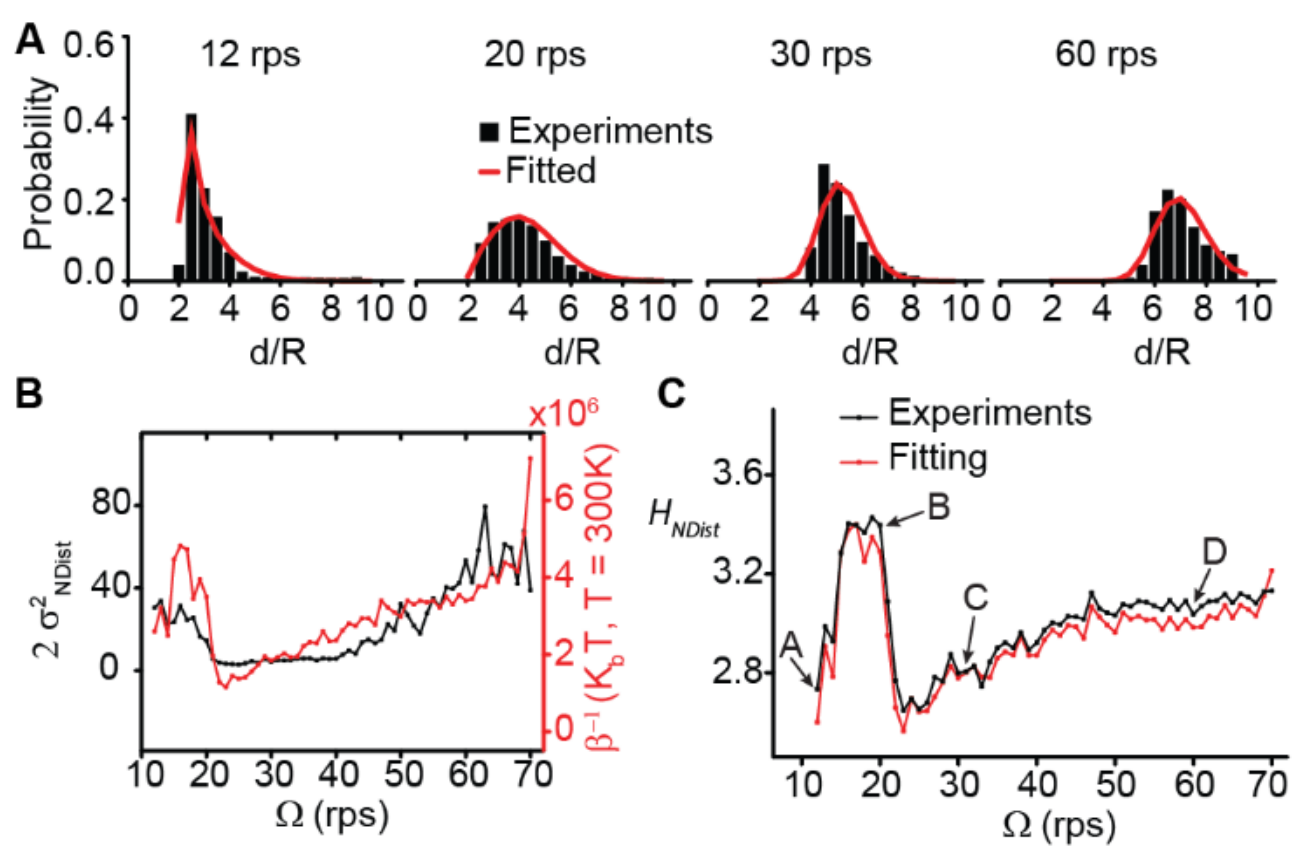

**Fig. 2 Neighbour distances as the variable for the calculation of the Shannon entropy to quantify the information content of the patterns**. (**A**) Representative experimental steady-state distribution of nearest neighbour distances at 12, 20, 30 and 60 rps and the fitted distribution using the effective Hamiltonian. (**B**) The variance of the neighbour distance (black line) and 1/β (red line) as a function of the rotation speed of the magnetic field **Ω**. (**C**) Experimental and fitted Shannon entropy as a function of the rotation speed of the magnetic field **Ω**. The entropies are calculated from the experimental steady-state distribution of neighbour distances and the fitted distribution using an effective Hamiltonian. Letters A-D correspond to the representative patterns shown in Fig. 1D.

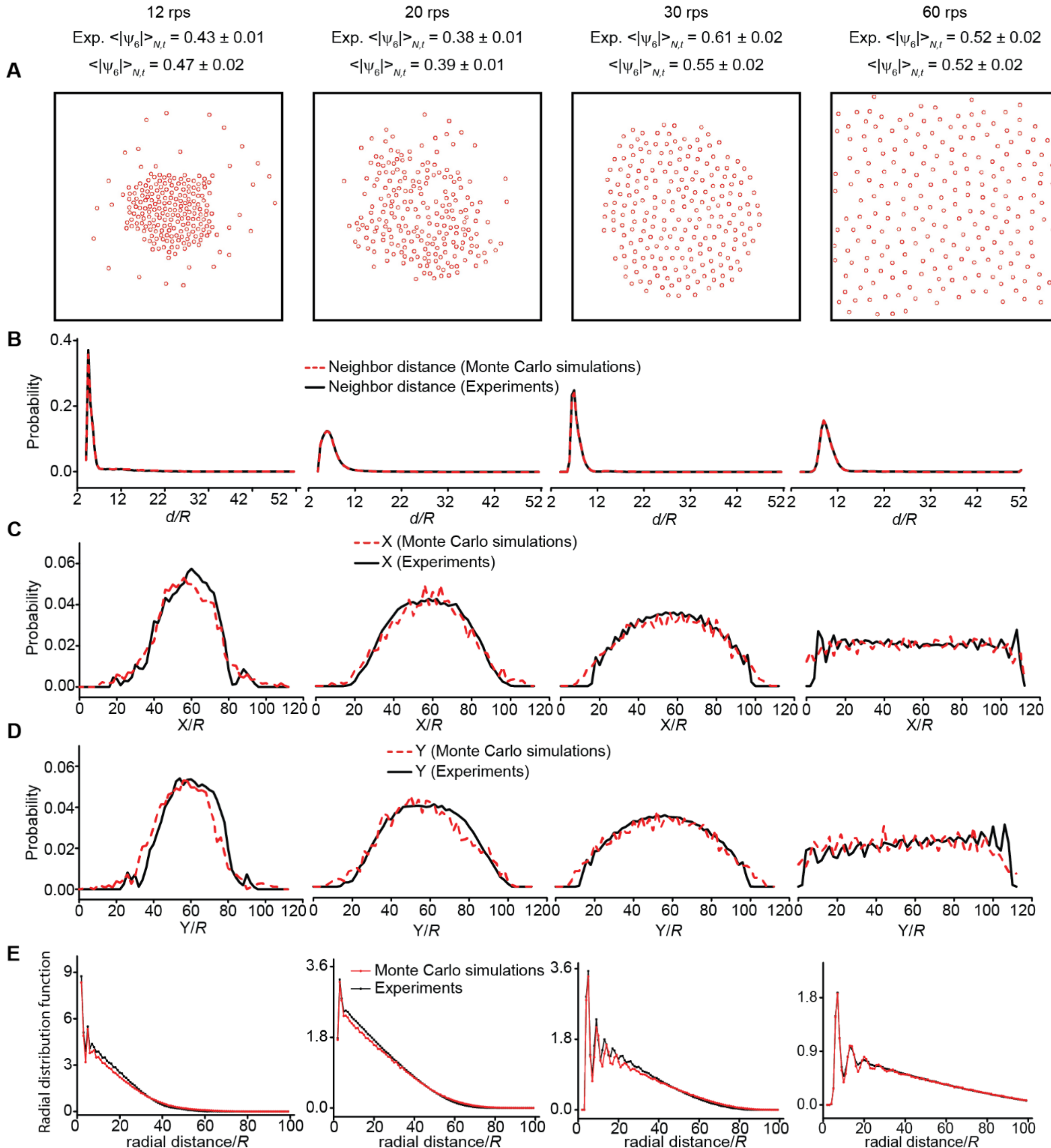

**Fig. 3 Reproducing disordered and ordered patterns using the distribution of neighbour distances only.** (**A**) Patterns produced from Monte Carlo simulations at representative rotation speeds. The means and the standard deviations of the simulated $\langle|\psi_6|\rangle_{N,t}$ are calculated from seven simulations. The means and the standard deviations of the experimental $\langle|\psi_6|\rangle_{N,t}$ are calculated from 75 – 1500 frames (1 – 20 seconds). (**B – D**) The corresponding distributions of neighbour distances, *x* coordinates and *y* coordinates from experiments and simulations. The distributions of *x* and *y* coordinates serve as one of the ways to compare the simulated patterns with those from the experiments. (**E**) Radial distribution functions of patterns from experiments and Monte Carlo simulations. The simulations data correspond to the average of 700 frames (100 frames × 7

simulations) at each rotation speed and the experimental data were averaged over 75 – 1500 frames (1 – 20 s).

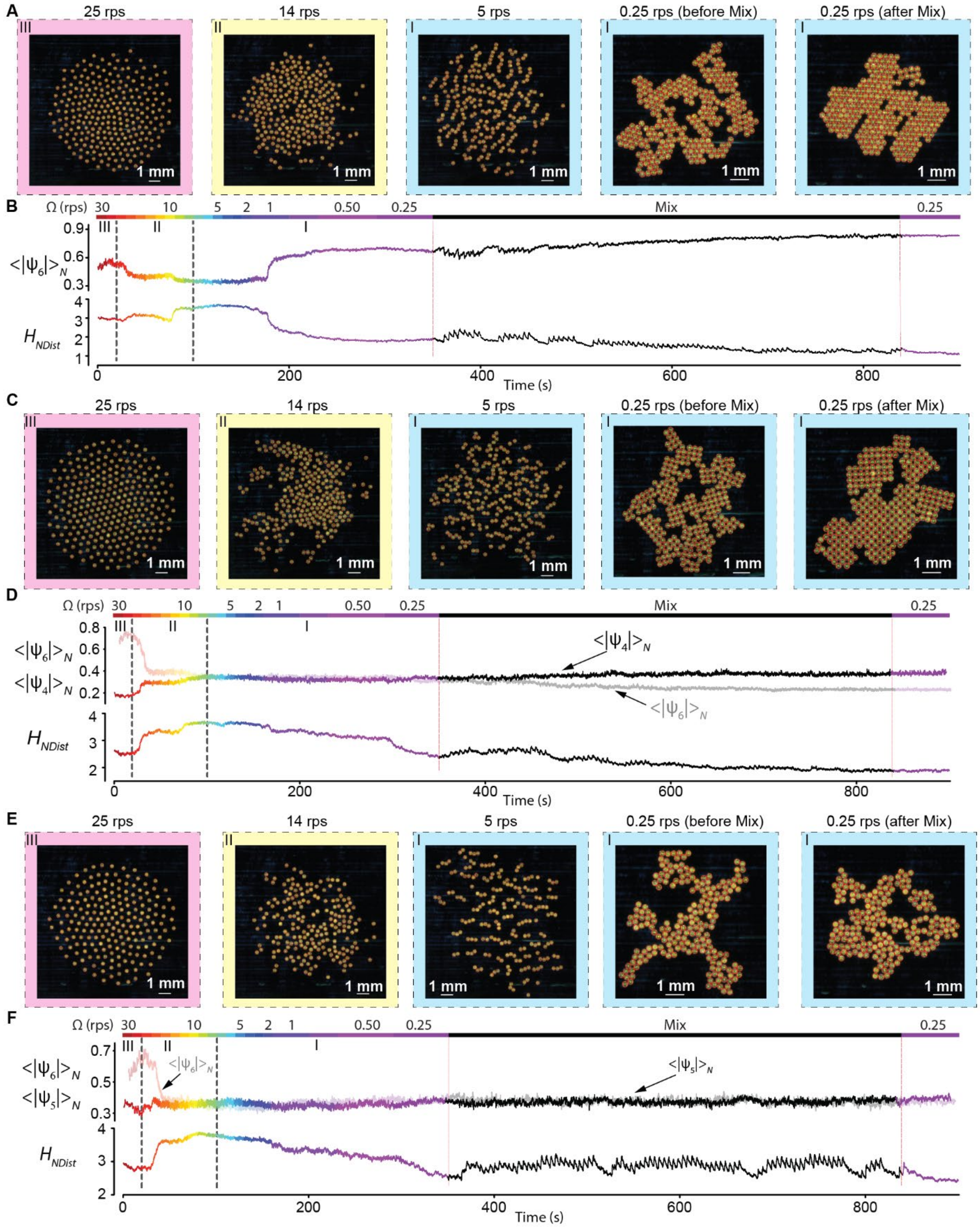

**Fig 4. Pattern transitions of hundreds of micro-disks to illustrate the relationship between information and order for the micro-disks with 4- , 5- and 6-fold symmetries. (A)**

Representative patterns of the micro-disks with 6-fold symmetry. The background colour corresponds to the regions I - III in the pairwise interactions in Fig. 1C. (**B**) The number averages of the norms of hexatic order parameters $\langle|\psi_6|\rangle_N$ and the entropies by neighbour distances $H_{NDist}$ as a function of time. The line colour indicates the rotation speed **Ω** of the applied magnetic field. "Mix" denotes low speeds at a low field strength mixed with high speeds at a high field strength. (**C**) Representative patterns of the micro-disks with 4-fold symmetry. (**D**) The number averages of the norms of hexatic order parameters $\langle|\psi_6|\rangle_N$, tetratic order parameters $\langle|\psi_4|\rangle_N$, and the entropies by neighbour distances $H_{NDist}$ as a function of time. (**E**) Representative patterns of the micro-disks with 5-fold symmetry. (**F**) The number averages of the norms of hexatic order parameters $\langle|\psi_6|\rangle_N$, pentatic order parameters $\langle|\psi_5|\rangle_N$, and the entropies by neighbour distances $H_{NDist}$ as a function of time. In the Mix region, the rotation speed and the strength of the external magnetic field were changed periodically. There are two types of mixtures, and both contain periods of 5 s. The first mixture consists of 1 s of 1 rps at 3 mT and 4 s of 0.25 rps at 0.5 mT. The second mixture consists of 1 s of 5 rps at 3 mT and 4 s of 0.25 rps at 0.5 mT. See the section on Detailed experimental protocols for full procedures. The Fourier spectra of the time series in the Mix region are in fig. S8.

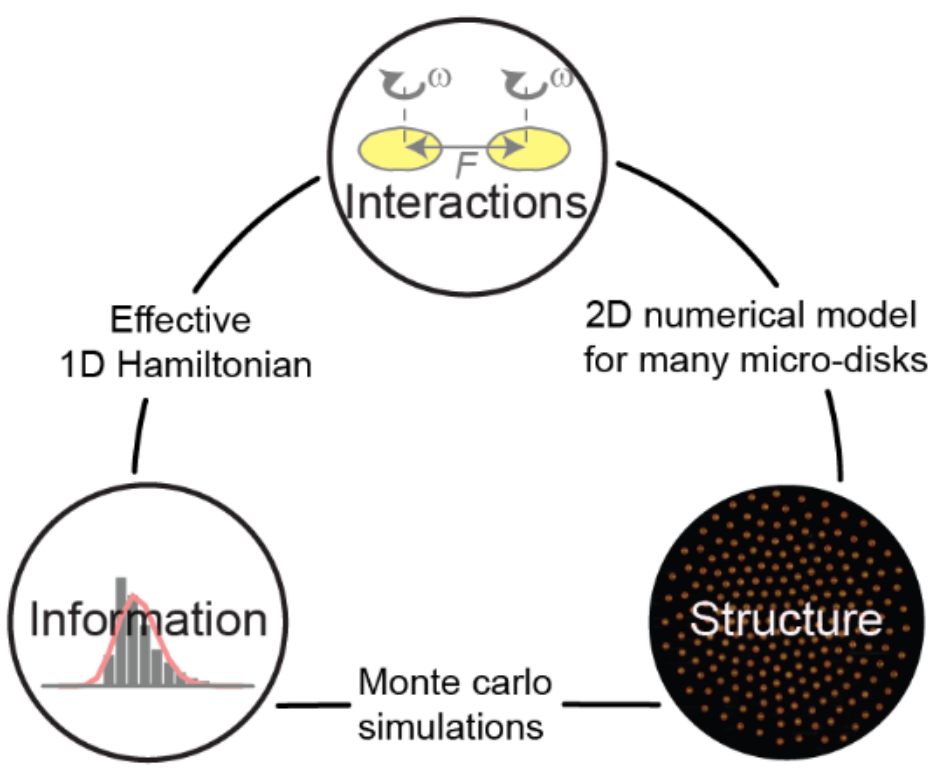

**Fig 5. The relations among information, structure, and interactions.**